\documentclass[aps,showpacs,prl,twocolumn]{revtex4-2}  

\usepackage{amsmath}
\usepackage{mathrsfs}
\usepackage{amssymb}
\usepackage[utf8]{inputenc}
\usepackage{graphicx,epstopdf,epsfig}
\usepackage{bm}
\usepackage{color}
\usepackage{times}
\usepackage{hyperref,url}

\newcommand{\etal}{{\emph{et al.}}} 		
\newcommand{\mr}[1]{{\mathrm{#1}}} 		
\newcommand{\bra}[1]{{\langle #1 \vert}} 		
\newcommand{\ket}[1]{{\vert #1 \rangle}} 		

\definecolor{clr}{rgb}{0,0.6,0.6}

\begin{document}

\title{Strong coupling and active cooling in a finite temperature hybrid atom-cavity system}
\author{L. F. Keary}
\author{J. D. Pritchard}
\email{jonathan.pritchard@strath.ac.uk}
\affiliation{Department of Physics, SUPA, Strathclyde University, Glasgow, G4 0NG, UK}

\date{\today}

\begin{abstract}
For a standard two-level atom coupled to the quantized field of a resonant cavity, finite temperature effects lead to thermal occupation of the cavity modes that obfuscates measurement of the quantum nature of the atom-light interaction. In this paper we demonstrate that using a hybrid system of a superconducting cavity coupled to a multi-level Rydberg atom it is possible to observe the quantum nature of strong coupling even at finite temperatures, and to exploit this coupling to permit cooling of the thermal microwave mode towards the ground-state, enabling observation of coherent vacuum Rabi oscillations even at 4~K for realistic experimental parameters. Cooling for multiple atoms is also explored, showing maximal cooling for small samples, making this a viable approach to cavity cooling with potential applications in long-range coupling of superconducting qubits via thermal waveguides.
\end{abstract}


\maketitle

Cavity quantum electrodynamics (QED) \cite{haroche06} reveal{\color{black}{s}} the quantum nature of the interaction between a harmonic oscillator and a two-level quantum system, resulting in a discrete level splitting and coherent transfer of excitation between the quantized field mode and the qubit. {\color{black}{Strong coupling has been reported in a diverse range of physical systems including optical \cite{thompson92}, microwave  \cite{bernardot92} and mechanical \cite{poot12,aspelmeyer14} oscillators coupled to atoms, spins and more recently artificial superconducting qubits \cite{wallraff04}, through observation of vacuum Rabi oscillations. Superconducting qubits}} are excellent candidates for scalable quantum information processing \cite{clarke08,devoret13,fowler12}, offering fast gate times \cite{barends14,martinis14} with state-of-the-art experiments already satisfying the required fault-tolerant thresholds \cite{fowler12,barends14}.

{\color{black}{Superconducting}} circuits are also ideal for realizing hybrid technologies \cite{xiang13,kurizki17,yiwen21}, permitting integration with micro-mechanical resonators for ground-state cooling \cite{oconnell10,teufel11} and opto-mechanical conversion of single photons from the optical to microwave domain \cite{andrews14,tian15} as well as coupling to solid state \cite{amsuss11,kubo11} or atomic qubits \cite{hatterman17, morgan20, kaiser21} for quantum memories. Highly excited Rydberg states provide an attractive candidate for this purpose, where the large electric dipole moment offers {\color{black}{single atom}} strong coupling which can be used for quantum gates \cite{petrosyan09,pritchard14}, long-distance entanglement of atomic ensembles via {\color{black}{microwave coupling}} \cite{petrosyan08,sarkany15} and high speed optical to microwave conversion rates \cite{gard17,covey19a}. 

\begin{figure}[t]
\includegraphics{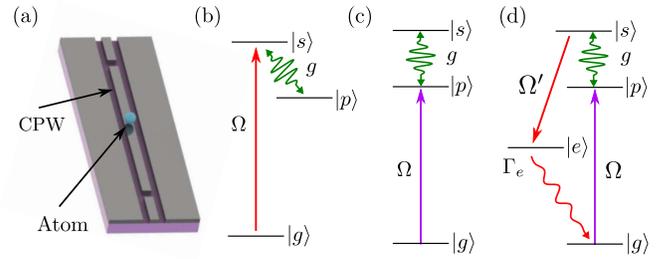}
\caption{(Color online). Hybrid Atom-Cavity Scheme (a) Atom strongly coupled to a superconducting microwave coplanar waveguide resonator (CPW) (b) Lambda level scheme ($\Lambda$) (c) Ladder level scheme ($\Xi$) (d) Cavity cooling scheme.}
\label{fig1}
\end{figure}

For low frequency mechanical and microwave oscillators finite temperature effects are significant due to thermal occupation of the cavity mode. This results in a transition in the cavity transmission spectrum from the characteristic vacuum Rabi splitting to that of an empty cavity, preventing observation of the quantum nature of the atom-light interaction \cite{fink10,henschel10}. {\color{black}{Direct observation of vacuum splitting can be recovered by cooling the system to mK temperatures inside a dilution refrigerator}}. If however the {\color{black}{ideal}} two-level system is replaced by a multi-level qubit, such as an atom with highly excited Rydberg levels, it is possible to overcome these limitations to achieve ground-state cooling of mechanical oscillators \cite{bariani14} to enable robust manipulation of quantum states even at finite temperature \cite{sarkany15}. Previous work has also demonstrated cavity cooling using flying qubits from an atomic beam traversing a microwave cavity mode~\cite{nogues99,raimond01,bernu08} and proposed cooling via spontaneous emission using a stationary ensemble of $\sim10^3$ Rydberg atoms~\cite{sarkany18}. 

{\color{black}{In this Letter we show that whilst finite temperature effects prevent observation of the coherent cavity dynamics such as vacuum Rabi oscillations, it is possible to recover the characteristic vacuum splitting expected for a cavity QED system through the use of multi-level Rydberg atoms. We observe vacuum Rabi splitting at temperatures as high as 4~K, essential for implementing robust long-distance entanglement \cite{sarkany15}}}. In this regime we explore the feasibility {\color{black}{of}} active cavity cooling using a single atom, going beyond asymptotic scaling laws used in earlier studies \cite{sarkany18}. {\color{black}{We perform}} a full simulation of the atom-cavity master equation and {\color{black}{consider}} the case of coupling the Rydberg state to a rapidly decaying intermediate level to increase {\color{black}{dissipation}}. In contrast to the predictions of \cite{sarkany18}, our cooling scheme results in a non-thermal steady-state cavity mode, with an enhanced population of the vacuum state combined with increased population of high-lying states. {\color{black}{For experimentally accessible}} parameters it is not possible to cool to the absolute ground-state of the cavity mode. {\color{black}{However, we demonstrate that active cooling using a single Rydberg atom is sufficient to observe atom-cavity dynamics without pre-cooling to sub~K temperatures typically required for circuit QED}}. This cooling scheme can be utilized to aid implementation of long-distance microwave communication to couple distant qubits via thermal waveguides by cooling the intermediate resonators \cite{xiang17,vermersch17}, or for demonstration of hybrid quantum computing.


The hybrid atom-cavity system is illustrated in Fig.~\ref{fig1}(a){\color{black}{. We}} consider a highly excited Rydberg atom strongly coupled to a superconducting microwave coplanar waveguide resonator (CPW) using the large electric dipole moment between nearby Rydberg states. For an atom initially in the ground state $\ket{g}$, population is transferred to a state coupled to the microwave cavity. We consider {\color{black}{two schemes}}, a lambda ($\Lambda$)-scheme using a two-photon excitation to couple $\ket{g}$ to an $\ket{s}$ state (shown in Fig.~\ref{fig1}(b)) and a ladder ($\Xi$) configuration using a single-photon UV transition to couple $\ket{g}$ to a $\ket{p}$ state (shown in Fig.~\ref{fig1}(c)).

Applying the rotating wave approximation for both the cavity and the two-photon transitions, the Hamiltonian associated with each configuration is given by 
\begin{equation}
\hat{\mathscr{H}}_x= \hat{\mathscr{H}}_c+\hat{\mathscr{H}}_a^x+\hat{\mathscr{H}}^x_\Omega+\hat{\mathscr{H}}_\mr{JC},
\end{equation}
where $x=\Lambda$ or $\Xi$, $\hat{\mathscr{H}}_c=\hbar\omega_c(\hat{a}^\dagger\hat{a}+1/2)$ is the energy operator for the microwave cavity mode with frequency $\omega_c$ and $\hat{a}$ ($\hat{a}^\dagger$) are the photon creation (annihilation) operators. The unperturbed atomic energy level operator is equal to
\begin{subequations}

\begin{eqnarray}
\hat{\mathscr{H}}_a^\Lambda &= - \hbar(\Delta^{\Lambda}+\omega_{a})\ket{p}\bra{p} - \hbar\Delta^{\Lambda} \ket{s}\bra{s},\\
\hat{\mathscr{H}}_a^\Xi &=  -\hbar\Delta^{\Xi}\ket{p}\bra{p} - \hbar(\Delta^{\Xi}-\omega_a)\ket{s}\bra{s},
\end{eqnarray}
\end{subequations}
{\color{black}{where we assume the cavity is resonant with the atomic transition frequency, $\omega_{c} =\omega_{a}= \omega_{s}- \omega_{p}$,}} and {\color{black}{$\Delta^{\Lambda,\Xi} = \omega-(\omega_{s,p}-\omega_g)$}} is the detuning from $\vert g \rangle$ to the respective Rydberg state for the chosen excitation scheme. The operator for the semi-classical coupling from $\ket{g}$ to {\color{black}{the}} Rydberg state is 
\begin{subequations}
\begin{eqnarray}
\hat{\mathscr{H}}_{\Omega}^{\Lambda} &=\hbar\Omega/2(\ket{s}\bra{g}+\ket{g}\bra{s}),\\
\hat{\mathscr{H}}_{\Omega}^{\Xi} &=\hbar\Omega/2(\ket{p}\bra{g}+\ket{g}\bra{p}),
\end{eqnarray}
\end{subequations}
where $\Omega$ is the Rabi frequency. Finally, the atom-cavity coupling is given by the Jaynes-Cummings operator, $\hat{\mathscr{H}}_\mr{JC} = \hbar g(\hat{a}\ket{s}\bra{p}+\hat{a}^\dagger\ket{p}\bra{s})$, with vacuum Rabi frequency $g=\mathcal{E}d/\hbar$, dependent upon the cavity electric field at the atom $\mathcal{E}$. For both configurations, $\ket{s}$ represents the excited state in the coherent atom-cavity coupling.
 
The strong coupling of the atom-cavity system results the in the formation of dressed eigenstates $\ket{\pm,n} = \ket{p,n} \pm \ket{s,n-1}$ split by energy $\pm\sqrt{n}g$ where $n$ is the number of excitations within the cavity mode. Observation of coherent dynamics requires that the cavity coupling dominates over effects of spontaneous decay from the atomic levels $\Gamma_{s,p}$ and decay rate of photons from the cavity mode $\kappa$, meaning the cooperativity $\mathcal{C}=g^2/\kappa\Gamma_{s,p}>1$. 
 
Thermal loading of the cavity mode leads to a thermal bath with mean photon number {\color{black}{$\bar{n}_{th}=[\exp(\hbar\omega_c/k_\mr{B}T)-1]^{-1}$ that incoherently drives excitations over a range of $\vert \pm,n \rangle$ with different Rabi frequencies.}} This dissipative process can be characterized by the Lindblad operator for the cavity field as
\begin{equation}
\begin{split}
\mathcal{L}_c =& (1+\bar{n}_\mathrm{th})\kappa(\hat{a}\rho\hat{a}^\dagger-\hat{a}^\dagger\hat{a}\rho/2-\rho\hat{a}^\dagger\hat{a}/2)\\
&+\bar{n}_\mathrm{th}\kappa(\hat{a}^\dagger\rho\hat{a}-\hat{a}\hat{a}^\dagger\rho/2-\rho\hat{a}\hat{a}^\dagger/2),
\end{split}
\label{eq:Lc}
\end{equation}
which describes the addition or removal of photons from the cavity as a result of the thermal environment. Additional Lindblad terms to account for spontaneous decay of the Rydberg states at rates $\Gamma_i$ are given by $\mathcal{L}_{i}=\sqrt{\Gamma_{i}}\ket{g}\bra{i}$ for $i=s,p$.  The decay rate of photons from the cavity resonator is determined by the quality factor $Q$ as $\kappa=\omega_c/Q$. 

For a 5~GHz resonator at $T=4~$K, the thermal occupation of $\bar{n}_\mr{th} \sim 16$ precludes {\color{black}{observation of}} strong coupling. A compromise can be made by increasing the cavity resonance frequency to $\omega_c/2\pi=15$~GHz to decrease $\bar{n}_\mathrm{th}\sim5$~photons by using Rydberg states with lower principal quantum numbers. {\color{black}{Use of lower lying Rydberg states results in reduced dipole moments and shorter lifetimes, but can still reach the single atom strong coupling regime.}} In this work we consider the transition from $\ket{s}=\vert 65S_{1/2},m_j=1/2\rangle\rightarrow {\color{black}{\ket{p}}}=\vert64P_{3/2},m_j=1/2\rangle$ in Cs at $\omega_a/2\pi=15$~GHz, with a dipole matrix element $d\sim\sqrt{2/9}\times3993~ea_0$ \cite{sibalic17} and {\color{black}{lifetimes $\tau_{s,p}$ of 289 and 689$~\mu$s, respectively}} \cite{beterov09}. The coupling strength is dependent upon the RMS voltage $V_0^\mathrm{rms}=\sqrt{\hbar\omega_c/2C}$ where $C$ is the total capacitance of the resonator \cite{goppl08}. For optimized resonators, the fundamental mode of a $\lambda/2$ resonator is $V_0^\mathrm{rms}\sim 6~\mu$V at 15~GHz, corresponding to an RMS electric field amplitude $\mathcal{E}_0=0.2$~V/m \cite{beck16} resulting in a coupling strength of $g/2\pi\sim4$~MHz. 

{\color{black}{We further consider the experimentally attainable $Q$ factors for the case of finite temperature CPW resonators. At temperatures below 100~mK, CPWs can achieve $Q\gtrsim 10^6$ \cite{megrant12}, however at elevated temperatures resonator losses are dominated by thermal quasiparticles, leading to a {\color{black}{significant}} reduction in resonator quality factor. One approach to suppressing losses is to engineer the CPW geometry, {\color{black}{Ref.~\cite{beck16} demonstrates a $Q\gtrsim3\times10^4$ for a 5~GHz Nb resonator at 4.2~K through suppression of kinetic inductance.}} To recover $Q$ at higher frequencies, it is necessary to utilise materials with higher critical temperatures such as NbN with $T_c=16$~K \cite{yoshida95}. For a 15 GHz resonator using a 500 nm NbN film thickness we estimate a theoretical $Q \sim 10^{5}$.}}

\begin{figure}[t]
\includegraphics{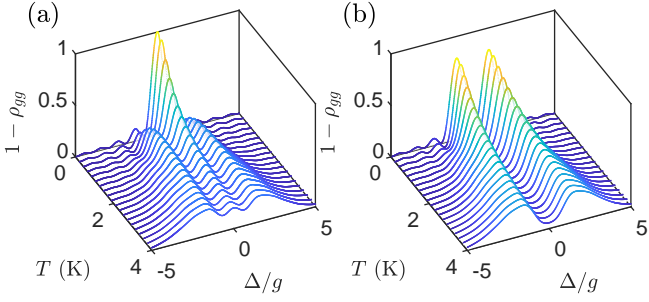}
\caption{(Color online). Ground state transfer as a function of $\Delta$ and $T$ for (a) $\Xi$- and (b) $\Lambda$-systems after a pulse duration $\tau=\pi/\Omega$ for $g/2\pi=4$~MHz, $\Omega=0.5g$ and $Q=10^5$.} 
\label{fig:thermal}
\end{figure}

{\color{black}{To demonstrate the impact of finite temperature effects we solve the master equation for the evolution of the system for a pulse }\color{black}{of duration $\tau = \pi/\Omega$}\color{black}{ applied to an atom initially in state $\ket{g}$ for $\Omega = 0.5 g$ as a function of temperature and detuning.}} Numerical simulations were performed in QuTiP \cite{johansson13} using the parameters above with $Q=10^5$. Results in Fig.~\ref{fig:thermal}, show state transfer out of $\ket{g}$ as a function of frequency. For the $\Xi$-system in Fig.~\ref{fig:thermal}(a) there is initially no signature of the atom-cavity coupling as the atom is excited to state $\ket{p,0}$ which does not undergo vacuum Rabi oscillations. As the temperature increases, the thermal loading of the resonator can be observed leading to broad normal-mode splitting either side of the central resonant feature. {\color{black}{For the $\Lambda$-system, at 0~K two high contrast peaks at $\Delta=\pm g$ arise from excitation of the $\ket{\pm,1}$ state. At increasing temperatures the peaks broaden and {\color{black}{reduce in}} amplitude due to the occupation of thermal modes with a splitting of $\pm\sqrt{n}g$}}. This leads to a drop in both contrast and resolution when performing spectroscopy at $T>0$, and shows that whilst finite temperature effects enable signatures of strong coupling in the $\Xi$-system that were not previously visible, the $\Lambda$-system offers superior contrast for characterizing the atom-cavity coupling.

\begin{figure}[t!]
\includegraphics[width=1\columnwidth]{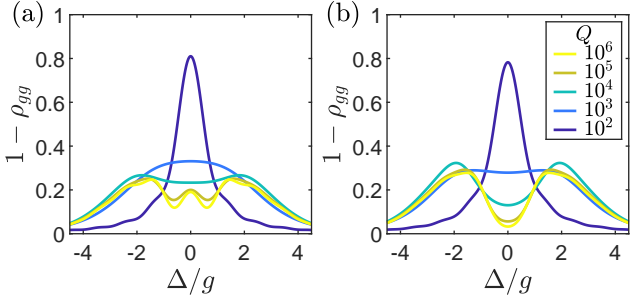}
\caption{(Color online).  Ground state transfer as a function of $Q$ for (a) $\Xi$ and (b) $\Lambda$-system for $g/2\pi=4$~MHz, $\Omega=0.5g$ and T = 4~K.
\label{fig:Q}}
\end{figure}

The requirement to achieve {\color{black}{strong coupling further restricts the cavity parameters}}. As the {\color{black}{Rydberg state}} linewidths $\Gamma_{s,p}\ll g$, the cooperativity $\mathcal{C}>1$ for a wide range of cavity $Q$. A better figure of merit is consideration of the average number of Rabi oscillations $n_\mathrm{Rabi}=2g/(\Gamma+\kappa)\sim{2g/\kappa}$ {\color{black}{requiring $\kappa< g$ for $n_\mathrm{Rabi}>1$}}. This limit is demonstrated in Fig.~\ref{fig:Q} which shows the results of the same spectroscopic measurement at 4~K for varying cavity $Q$ factors, further highlighting the resilience of the $\Lambda$-system for performing experiments at finite temperature where normal mode splitting can be observed for $Q>10^3$ although for the $\Xi$-system even at $Q=10^4$ the splitting is poorly resolved.

Whilst these results identify experimentally viable parameters in which it is possible to demonstrate strong-coupling for a finite temperature system, we can further exploit the multi-level nature of our atomic qubit to implement active cavity cooling by combining our two excitation pathways using the level scheme illustrated in Fig.~\ref{fig1}(d). Previous work considered the effect of cooling directly via spontaneous emission from the Rydberg state alone using a simple rate equation model \cite{sarkany18}. Here we explicitly model the full system evolution {\color{black}{where the Rydberg state is coupled}} to a low-lying excited state $\ket{e}$ with a decay linewidth {\color{black}{$\Gamma_e\gg\Gamma_{s,p}$}} to enhance the spontaneous emission rate. In this system cooling is achieved by continuously driving from $\ket{g,n}\rightarrow\ket{p,n}$, where the atom can absorb a photon from the cavity to reach state $\ket{s,n-1}$ which is then coupled to $\ket{e,n-1}$ by an additional driving laser with Rabi frequency $\Omega'$. This enables rapid dissipation via spontaneous emission to reach $\ket{g,n-1}$, reducing the thermal occupation in the cavity.

The Hamiltonian for the four-level cooling system is given by
\begin{equation}
\hat{\mathscr{H}}_\mr{cool} = \hat{\mathscr{H}}_\Xi-\hbar(\Delta-\Delta'-\omega_{a}) \ket{e}\bra{e} +\hbar\Omega'/2(\ket{e}\bra{s}+\ket{s}\bra{e}),
\end{equation}
where $\Delta' = \omega'-(\omega_s-\omega_e)$ is the detuning of the laser coupling $\ket{e}\rightarrow\ket{s}$, and we explicitly consider the case $\ket{e}=6P_{3/2}$ with $\Gamma_e/2\pi = 5.2$~MHz which gives rise to the additional Lindblad operator $\mathcal{L}_{e}=\sqrt{\Gamma_{e}}\ket{g}\bra{e}$.

\begin{figure}[t]
\includegraphics[width=1\columnwidth]{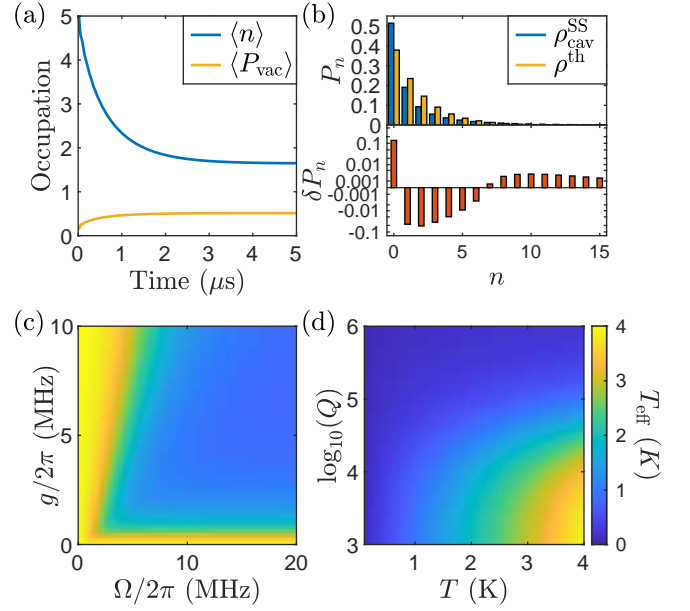}
\caption{(Color online). Cavity cooling (a) Temporal evolution for cooling scheme in Fig.~\ref{fig1}(d) using $\Omega=\Omega'=2\pi\times10$~MHz, $g/2\pi = 4$~MHz, $Q = 10^{5}$, $T = 4$~K (b) Steady state solution vs thermal field distribution (c) Steady-state effective temperature $T_\mr{eff}$ as a function of $g$ and $\Omega'=\Omega$ (d) $T_\mr{eff}$  as a function of $Q$ and $T$. \label{fig:fig4}}
\end{figure}

We model the cavity evolution for the case of resonant excitation with $\Delta=\Delta'=0$, $\Omega = \Omega' = 2\pi\times10$~MHz and $Q=10^5$ for a cavity at $T=4$~K, with results in Fig.~\ref{fig:fig4}(a) showing rapid reduction of the average photon number to a steady-state value of $\langle n \rangle^\mathrm{SS} = 1.65$ within a few $\mu$s and increased occupation of the {\color{black}{cavity vacuum state $\langle P_\mr{vac} \rangle= \operatorname{Tr}(\ket{0}\bra{0}\rho)$}}. Following from Bariani \etal{} \cite{bariani14} we define an effective temperature from the steady-state vacuum mode $\langle P_\mr{vac} \rangle^\mathrm{SS}$ equal to 
\begin{equation}
T_\mr{eff}  =-\frac{\hbar {\color{black}{\omega_{c}}}}{k_{B} \log{(1-\langle P_\mr{vac} \rangle^\mathrm{SS})}}, 
\label{Eq:Teff}
\end{equation}
resulting in a steady-state temperature of $T_\mr{eff}\sim1$~K for the parameters considered. The expression for $T_\mr{eff}$ given here provides a metric to quantify the system temperature since cavity cooling results in a {\color{black}{steady-state field}} which is not well described as a thermal distribution. This is highlighted in Fig.~\ref{fig:fig4}(b) where we compare the steady-state cavity field to a thermal field with the same mean photon number. {\color{black}{This reveals that whilst cavity cooling enhances the ground-state fraction , it leaves a residual occupation in high lying states with $n\gtrsim5$ which are split by $\sqrt{n}g>\Omega$ and thus are outside the linewidth of the excitation, giving an extended tail to the photon number distribution and preventing cooling to the ground-state of the cavity mode.}} {\color{black}{This observation implies cooling should be enhanced by increasing $\Omega$ to address {\color{black}{higher photon}} states, as verified in Fig.~\ref{fig:fig4}(c) which plots the steady-state $T_\mr{eff}$ as a function of $g$ and $\Omega$. }}

In Fig.~\ref{fig:fig4}(d), we {\color{black}{present}} $T_\mr{eff}$ as a function of cavity $Q$ and temperature for $\Omega=\Omega'=2\pi\times10$~MHz. This shows that for a system at 4~K, the minimum cavity quality factor required for effective cooling to $T<1$~K is $Q \geq 10^{5}$, placing a tighter constraint on the required cavity $Q$ than {\color{black}{is}} needed to observe signatures of strong-coupling. 

\begin{figure}[t!]
\includegraphics{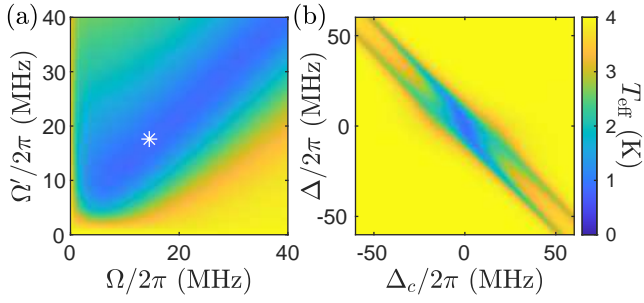}
\caption{(Color online). Cavity cooling optimization for $g/2\pi=4$~MHz, $Q=10^5$ and $T=4$~K (a) Optimal cooling gives $\Omega/2\pi = 14.5$~MHz, $\Omega'/2\pi=17.6$~MHz for $\Delta=\Delta'=0$ (b) Effective temperature as a function of cavity detuning $\Delta_c$ and $\Delta$ for $\Delta'=0$ showing maximum cooling efficiency occurs for $\Delta=\Delta_c=0$.\label{fig:app}}
\end{figure}

We perform a further optimization of the cooling parameters for the case of $T=4$~K to minimize $T_\mr{eff}$ as shown in Fig.~\ref{fig:app}. This yields $\Omega/2\pi =14.5$~MHz, $\Omega'/2\pi =17.6$~MHz, {\color{black}{with a mean photon number of $\langle n \rangle^\mathrm{SS} = 1.14$}} and $T_\mr{eff} = 0.87$~K. Contrary to the expectation based on the proposal in Ref.~\cite{sarkany18} for cooling via spontaneous decay of the Rydberg state, we obtained no improvement in cavity cooling when operating at a non-zero $\Delta$ or $\Delta_{c}$ as {\color{black}{shown}} in Fig.~\ref{fig:app}(b).  
 
\begin{figure}[t]
\includegraphics[width=1\columnwidth]{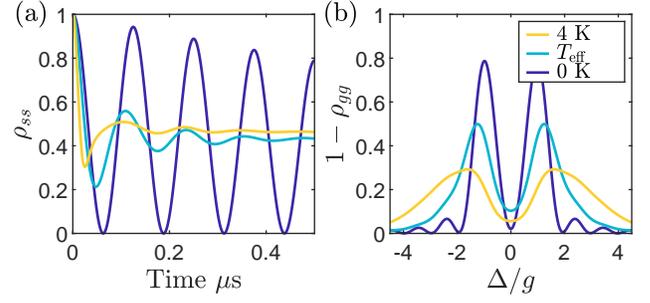}
\caption{(Color online). (a) Vacuum Rabi oscillation of state $\vert s \rangle$ and (b) Ground state transfer for the $\Lambda$-system for a cavity at 0~K, 4~K, and 4~K after active cooling {\color{black}{to $T_\mr{eff} = 0.87$~K}}. \label{fig:fig5}}
\end{figure}

The proposed cooling scheme thus provides a route to suppressed thermal occupation even for an initial temperature of $T=4$~K, which in turn should improve the coherence time for the atom-cavity interaction due to suppression of thermal driving via the cavity mode. To demonstrate this effect, we model the evolution of an atom initially prepared in state $\ket{s}$ for a cavity at 0~K, 4~K and for our 4~K cavity cooled to the steady-state using the optimized parameters above. Fig.~\ref{fig:fig5}(a) shows the resulting vacuum Rabi oscillations, with high contrast fringes observed for $T=0$~K, but strongly damped evolution with no discernible Rabi frequency in the case of a thermal cavity mode. {\color{black}{For the pre-cooled cavity however, we recover a coherent Rabi oscillation around the expected frequency with a damping time $\tau\gg\kappa$ due to the residual occupation of higher order photon modes}}. We further demonstrate the enhancement in characterizing the atom-cavity interaction by repeating the spectroscopy for the $\Lambda$-system described above for the case of the cooled cavity. Fig.~\ref{fig:fig5}(b) reveals this cooling significantly improves both the contrast and the linewidth of the resulting normal-mode splitting, greatly increasing the visibility of the feature of interest.

\begin{figure}[t!]
\includegraphics[width=1\columnwidth]{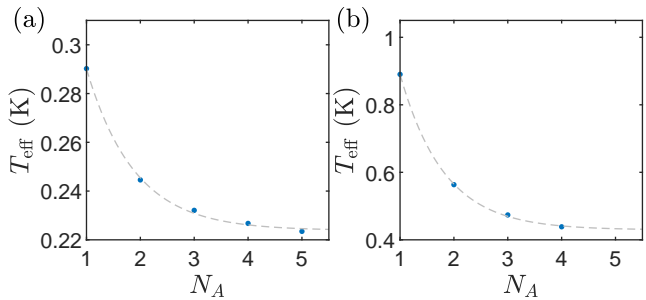}
\caption{(Color online). {\color{black}{Steady-state effective temperature from active cavity cooling vs atom number for an initial cavity temperature of}} (a) $T=1$~K and (b) $T=4$~K.\label{fig:fig6}}
\end{figure}

{\color{black}{Currently the cavity cooling is limited by finite $g$ and $\Gamma_{e}$, which act as a bottleneck in the cooling cycle. One approach to improving the cooling efficiency is to add more atoms to increase the rate at which thermal photons are absorbed.}} To model this scenario, we consider $N_{A}$ atoms coupled simultaneously with the same cavity field and neglect both collective spontaneous emission and Rydberg dipole-dipole interactions under the assumption that the atomic spacing $R \gg \lambda_{eg}$, such that atoms act independently {\color{black}{of each other}}. {\color{black}{This could be achieved using a dilute cloud of atoms \cite{kaiser21} or using single atoms trapped in optical tweezers \cite{schymik21}}}. The results of the multi-atom cavity cooling model are shown in Fig.~\ref{fig:fig6} for the optimized parameters using a cavity at {\color{black}{an initial}} temperature of (a) $T = 1$~K and (b) 4~K. We fit the {\color{black}{effective temperature as a function of atom number}} with phenomenological exponentially decaying function $T_\mr{eff} = T_\mr{eff}^\infty+\alpha\exp{(-N_{A}/N_c)}$ giving parameters $T_\mr{eff}^\infty= 0.22 (0.43)$~K, $\alpha = 0.20 (1.56)$~K and $N_c = 0.81 (0.89)$ for $T= 1 (4)$~K respectively. {\color{black}{At 1~K the enhancement associated with the multi-atom cooling scheme is marginal due to the cooling already operating in an efficient regime for a single atom, but for 4~K the enhancement is significant such that for $N_A\sim 5$ atoms the fit predicts $T_\mr{eff} = 430$~mK}}. In both cases, the improvement in the ground-state occupation saturates with relatively few atoms making implementation feasible in an experimental platform without requiring large atomic ensembles.  

In conclusion, we have investigated finite temperature effects in a hybrid atom-cavity system, where we identify experimentally realizable parameters for coupling a $\ket{s = 65S} \leftrightarrow \ket{p = 64P}$ Rydberg state to a 15~GHz superconducting coplanar microwave resonator. Our results show the impact of thermal excitations in suppressing the observable signatures of strong coupling due to broadening of the atomic resonances, and we determine a quality factor of $Q\ge10^4$ is required to recover a resolvable normal-mode splitting at 4~K. Furthermore, we demonstrate that for $Q\ge10^5$ active-cooling of the microwave mode is feasible by coupling the excited Rydberg level to a rapidly decaying intermediate state, achieving cooling to effective temperatures from 4~K to below 1 K for a single atom and 430~mK for $N_A\sim 5$ atoms. For a lower initial temperature the cooling cycle is more efficient, providing access to sub-K effective field temperatures without use of a dilution refrigerator and significantly increasing the visibility and coherence of quantized atom-light interactions. Whilst the required $Q\ge10^5$ at 15~GHz is challenging at 4~K, combining high-$T_c$ superconducting films with geometric engineering to suppress quasi-particle losses \cite{beck16} mean this should be experimentally feasible. These results show that hybrid systems offer a viable route to both realization of strong-coupling and active cooling even at finite temperatures, essential for implementing long-distance entanglement using the microwave cavity mode \cite{sarkany15}.

\begin{acknowledgments}
We thank M. Saffman and J. Jeffers for useful discussions. This work is supported by the EPSRC (Grant No. EP/N003527/1) and and through the EPSRC Centre for Doctoral Training in Delivering Quantum Technologies (Grant No. EP/P510270/1). The data presented in the paper are available here \cite{keary21data}.
\end{acknowledgments}
%

\end{document}